\documentclass[aps,prl,twocolumn,showpacs,groupedaddress,preprintnumbers]{revtex4}

\usepackage{graphicx}

\begin{document}

%
\preprint{MIFP-07-32}

\title{Proton Decay and  Flavor Violating Thresholds in SO(10) Models}


\author{Bhaskar Dutta}
\author{Yukihiro Mimura}
\author{Rabindra N. Mohapatra$^\dagger$}
\affiliation{
Department of Physics, Texas A\&M University,
College Station, TX 77843-4242, USA
}
\affiliation{
$^\dagger$Department of Physics, University of Maryland,
College Park, MD 20742, USA
}


\date{\today}

\begin{abstract}
Discovery of neutrino mass has put the spotlight on supersymmetric SO(10)
as a natural candidate for grand unification of forces and matter.
However, the suppression of proton decay is a major problem
in any supersymmetric grand unified models.
In this paper we show how to alleviate this problem
by simple threshold effect which raises the colored Higgsino masses
and the grand unification scale to $\gtrsim 10^{17}$ GeV.
There exist only four types of 
fields arising from different SO(10) representations which can generate
this kind of threshold effects.
Some of these fields also generate  a sizable flavor violation
in the quark sector
compared to the lepton sector.
The $b$-$\tau$ unification can work in these types of
models even for intermediate values of $\tan\beta$.
\end{abstract}

\pacs{12.10.Dm, 12.10.Kt, 12.60.Jv, 12.15.Ff}

\maketitle


Supersymmetry (SUSY)
is an attractive and widely discussed candidate for physics
beyond the standard model (SM) at the TeV scale. In addition to solving
the gauge hierarchy problem, it has the appealing feature of leading to
 unification of three SM gauge
couplings at a high scale of around $2\times 10^{16}$ GeV if there is no
new physics between the TeV and grand unified theory (GUT) scale. The coupling unification would
suggest that the SM gauge groups are grand unified into one simple group.
 An additional boost to the argument for
grand unification comes from the understanding of small neutrino masses
via the seesaw mechanism, which suggests the breaking of B-L symmetry
around the GUT scale. The simplest  GUT that
embodies all these features including the seesaw mechanism is SUSY SO(10)
group, where all quarks and leptons 
are unified in one $\bf 16$-dimensional spinor representation.

The downside of quark-lepton unification is that it
predicts an unstable proton. Searches for proton decay have so far have
not been successful and put a lower limit on its lifetime of $\agt
10^{33}$ years. Clearly, the observation of nucleon
decay would provide the most direct test of the GUT models.
However, the present lower bound on the nucleon lifetime
 imposes severe constraints
on the SUSY GUT models \cite{Hisano:2000dg,Babu:1997js},
if SUSY particles are less than 2-3 TeV.
The main sources of these constraints are the dimension five operators
induced by colored Higgsino fields which accompany
 the SM Higgs in the GUT theory \cite{Sakai:1981pk}.
The Higgsino couplings are then related to the fermion masses and are
less free for adjustment. One way to escape these constraints is to add
additional
symmetries above and beyond SO(10), which eliminates such dimension five
operators altogether.
In this study, we assume that the dimension five operators
do really exist and point out a new
way to suppress their effects.
%

As is well known, in the presence of the Higgsino induced dimension five
operators, the nucleon lifetime depends on
SUSY particles' masses, colored Higgsino mass (chosen to be of order of
the GUT scale),
and Yukawa texture for fermions. In a recent paper, we pointed out that
one way to suppress these contributions is to use suitable Yukawa texture
 \cite{Dutta:2004zh} and presented an  SO(10) example that uses only
renormalizable couplings for fermion masses. This suppresses proton
decay even for
the color triplet Higgsino mass at $2\times 10^{16}$ GeV.
Such a structure provides hierarchical mass for quarks and leptons, small
$U_{e3}$ (one of the neutrino mixings) and
 large atmospheric and solar neutrino mixings. Here we consider an
alternative approach where we do not constrain the Yukawa texture;
instead we introduce new thresholds to suppress
nucleon decay while at the same time maintaining gauge coupling
unification.
 %

This approach is nontrivial due to constraints of gauge unification.
For example, in the minimal SU(5) model, coupling unification imposes a
very stringent upper bound on the colored Higgsino mass making it
impossible to
satisfy current experimental bounds of the nucleon lifetime
\cite{Hisano:2000dg}.
On the other hand, the rich multiplet structure of SO(10) allows new
sub-GUT scale thresholds even in its minimal version. Of course it is
apriori not clear whether coupling
unification will necessarily impose constraints on Higgsino
mass when one includes new thresholds. We explore this issue here.

An interesting
point about new thresholds is that they may be accessible at low energies
via new flavor effects. The point is that
in SUSY models, the flavor degeneracy of  squarks
and sleptons are often assumed at high scale to avoid
flavor changing neutral currents (FCNCs). However loop corrections in the
presence of new thresholds can induce flavor violation and can provide a
way to test indirectly for GUT scale particle spectrum.
In the MSSM, the induced FCNCs in the quark sector from the evolution of
renormalization group
equations (RGE)
are small due to the small quark mixings whereas
in the lepton sector,
sizable effects can arise from the presence of right-handed neutrino
thresholds in the seesaw framework \cite{Borzumati:1986qx}.
New thresholds can make their presence felt both in the quark and the
lepton sector by altering these FCNC footprints in interesting ways.

In this Letter,
we will show that there exists a simple threshold
effect that makes the colored Higgsino mass and the
symmetry breaking scale larger
in SO(10) models,
and thus naturally suppressing proton decay.
In such a SO(10) breaking vacuum,
it is possible
that a sizable flavor violation is generated
in the quark sector
rather than the lepton sector.
We also emphasize that
investigating flavor violation in both quark and lepton
sectors 
 is important to select possible scenarios of GUT models and symmetry
breaking vacua.

At first, we will briefly examine 
the SO(10) GUT model buildings.
In order to break SO(10) symmetry down to the SM
gauge symmetry,
we employ
$\bf 210$ 
(four-antisymmetric tensor)
and
 ${\bf 126}(\Delta) + \overline{\bf 126}(\bar \Delta)$ representations.
In this choice, the SO(10) breaking vacua down to SM
can be dictated in the minimal number of parameters \cite{Aulakh:2003kg}.
The Higgs representations above can be replaced to
${\bf 45}+{\bf 54}$ and
${\bf 16} + \overline{\bf 16}$, respectively.
The Higgs spectrum and the breaking patterns of SO(10)
can be found in the Ref.\cite{Fukuyama:2004xs}.
One can also employ only a pair of vector-spinor representations
(${\bf 144} + \overline{\bf 144}$)
to break SO(10) down to SM at a single scale \cite{Babu:2005gx}.
All fermions are unified in $\bf 16$ representation $\psi_i$ in each generation $(i = 1,2,3)$.
The Higgs fields which couple to fermions in renormalizable Yukawa terms
are ${\bf 10}$ $(H)$, $\overline{\bf 126}$ $(\bar \Delta)$ and ${\bf 120}$ $(D)$:
\begin{equation}
W_Y = \frac12 h_{ij} \psi_i \psi_j H +\frac12 f_{ij} \psi_i \psi_j \bar \Delta
 +\frac12 h_{ij}^\prime \psi_i \psi_j D .
\end{equation}
The SO(10) invariance implies that the coupling matrices
$h$ and $f$ are symmetric and $h^\prime$ is antisymmetric.

The Yukawa couplings for quarks and leptons can be written as
\begin{eqnarray}
Y_u = \bar h + r_2 \bar f + r_3 \bar h^\prime,\ &&
Y_d = r_1 (\bar h + \bar f + \bar h^\prime), \\
Y_\nu = \bar h -3 r_2 \bar f + c_\nu \bar h^\prime,\ &&
Y_e = r_1 (\bar h - 3 \bar f + c_e \bar h^\prime),
\end{eqnarray}
where $u,d,e,\nu$ denote up-type quark, down-type quark,
charged-lepton, Dirac neutrino Yukawa couplings, respectively.
Notations such as $r_{1,2,3}$,
which are the functions of Higgs mixings,
 and $\bar h$, which is an original coupling $h$ multiplied by a Higgs mixing,
are given in the Ref.\cite{Dutta:2004zh}.
The parameter $r_1$ provides a freedom of $\tan\beta$
(ratio of vacuum expectation values (VEVs) for up and down Higgs doublets in MSSM).
When there is an exchange symmetry between $\Delta$ and $\bar \Delta$,
$r_1$ turns out to be 1 and then $\tan\beta$ is $\sim 50$.
If there is no such exchange symmetry,
$\tan\beta$ is a free parameter in the model
($t$-$b$-$\tau$ Yukawa unification is not satisfied),
while $b$-$\tau$ Yukawa unification
is still realized
approximately if $\bar f$ is small,
which is natural to obtain Georgi-Jarskog relation,
$3 m_s/m_b \simeq m_\mu/m_\tau$ at GUT scale.

There has been efforts to fit all fermion masses and mixings
using only $h$ and $f$ couplings (without $\bf 120$ Higgs) \cite{Babu:1992ia}.
However, such a minimal situation is disfavored \cite{Bajc:2005qe}.
In the minimal choice,
one needs fine-tuning to fit electron mass
since electron mass also becomes three times larger
than down quark mass naively.
In that case, the first and second generation component of
the coupling matrices are large,
and the proton decay suppression becomes really unnatural \cite{Dutta:2004zh}.
Even for the numerical fits of masses and mixings,
the fits are excluded more than 3 sigma level
if the minimal choice of Higgs content is considered~\cite{Bajc:2005qe}.
This is because
that
(1) $r_2$ is determined by the SO(10) breaking vacua
in the minimal choice.
(2) SO(10) breaking pattern is limited to obtain
proper neutrino mass scale.
(3) If the restricted SO(10) breaking vacuum is chosen, 
the gauge coupling unification does not occur.
However, such a disaster can be avoided if
we introduce additional Higgs fields as done in Refs.~\cite{Dutta:2004zh,Bertolini:2004eq}.
In these new scenarios, the constraint on  $r_2$ from the fermion fit is relaxed.
Therefore, in this Letter,
we do not pay much attention to the detail fitting of the fermion masses and mixings,
including neutrino mass scale since the fit is not restricted to the choice of a SO(10) breaking vacuum.

The dimension five operators ($LLLL$ and $RRRR$ operators) induced by colored Higgsino
are given as
\begin{equation}
-W_5 =\frac12 C_L^{ijkl} q_k q_l q_i \ell_j + C_R^{ijkl} e_k^c u_l^c u_i^c d_j^c .
\end{equation}
In the minimal SU(5) GUT,
$C_L$ and $C_R$ can be written by fermion Yukawa couplings, $Y_u$, $Y_d$ (or $Y_e$)
and there is no freedom to cancel.
In a SO(10) model, however,
$C_L$ and $C_R$ are written by combinations of $h$, $f$, and $h^\prime$
multiplied by colored Higgs mixings.
Therefore, there is a freedom to cancel (even in the minimal model) \cite{Goh:2003nv}.
However, such cancelation is quite unnatural if we make the general fitting
of fermion masses and mixings
since we have to introduce cancelation for each nucleon decay mode.
Actually, the coefficients of $f$ coupling in the $C_L$ and $C_R$
are opposite due to $D$-parity of the SO(10) symmetry,
and it is hard to suppress both $C_L$ and $C_R$
and only small $\tan\beta \sim 2$ remains available for the solutions
to satisfy current bounds.
We then need to introduce  suitable structures for  $h$, $f$ and $h^\prime$ couplings
to suppress the operators naturally \cite{Dutta:2004zh}.

As stated, our proposal is to suppress the proton decay by increasing the
mass of color-triplet Higgsinos since the dimension five operators
are generated by integrating out these particles.
We now study the condition under which this is possible consistent with
unification.

The colored Higgsino mass is bounded due to the
gauge coupling unification conditions.
The unification of the three gauge couplings
provides two independent relations on the particle mass spectrum
below the symmetry breaking scale \cite{Hisano:1992mh}.
The lightest colored Higgsino mass $M_{H_C}$ and $X$, $Y$ super heavy
gauge boson
mass $M_X$ are restricted by the following relations at 1-loop level:
\begin{eqnarray}
&&-2 \alpha_3^{-1}(m_Z)
+3\alpha_2^{-1}(m_Z)
-\alpha_1^{-1}(m_Z) \\
&&=
\frac{1}{2\pi} \left(\frac{12}5 \ln \frac{M_{H_C}}{m_Z}
+ \sum_I N_A^I \ln \frac{M_I}{\Lambda}
- 2 \ln \frac{m_{\rm SUSY}}{m_Z} \right),  \nonumber \\
&&-2\alpha_3^{-1}(m_Z)
-3\alpha_2^{-1}(m_Z)
+5\alpha_1^{-1}(m_Z) \\
&&=
\frac{1}{2\pi} \left(12 \ln \frac{M_X^2 \Lambda}{m_Z^3}
+ \sum_I N_B^I \ln \frac{M_I}{\Lambda}
+8 \ln \frac{m_{\rm SUSY}}{m_Z} \right),  \nonumber
\end{eqnarray}
where $N_A^I = 2 T_3(\phi_I) - 3 T_2 (\phi_I) + T_1 (\phi_I)$
and
$N_B^I = 2 T_3(\phi_I) + 3 T_2 (\phi_I) -5 T_1 (\phi_I)$
for SM decomposed fields $\phi_I$ (GUT particles except for
the lightest colored Higgs and $X$, $Y$ gauge bosons)
with mass $M_I$. $\Lambda$ is the GUT scale.
The Dynkin indices $T_{3,2,1}$ are given for $SU(3)_c \times SU(2)_L
\times U(1)_Y$
with canonical $U(1)_Y$ normalization by factor $3/5$.
We assume a single scale threshold $m_{\rm SUSY}$ for SUSY particle, just for
simplicity
to describe.
The GUT scale $\Lambda$ is cancelled if we take into account
the full multiplets in the equations.
For instance,
in the minimal SU(5) model,
additional Higgs fields are $({\bf 8},{\bf 1},0)$
and $({\bf 1},{\bf 3},0)$,
and $M_{({\bf 8},{\bf 1},0)} = M_{({\bf 1},{\bf 3},0)}$
is satisfied,
and then $\sum_I N_A^I \ln M_I/\Lambda = 0$ and
$\sum_I N_B^I \ln M_I/\Lambda = 12 \ln M_{({\bf 8},{\bf 1},0)}/\Lambda$.
As a result,
$M_{H_C}$ and $M_X^2 M_{({\bf 8},{\bf 1},0)}$ are constrained
from the measurements of the gauge couplings,
and the bound of colored Higgsino mass is
$M_{H_C} \leq 3.6 \times 10^{15}$ GeV \cite{Hisano:2000dg,Hisano:1992mh}.

The lightest colored Higgs mass scale are always comparable to (or
smaller than)
the heavy gauge boson masses
since it depends on the SO(10) breaking VEVs.
Therefore,
to increase the $M_{H_C}$ 
 bound,
we need negative contribution for both $\sum_I N_A^I \ln M_I/\Lambda$
and $\sum_I N_B^I \ln M_I/\Lambda$.
To realize such a situation,
we need a light field whose $N_A$ and $N_B$ are both positive.
If this field splits from the other SM decomposed fields and become
light for a given vacuum,
$M_{H_C}$ can be larger.
%
We list such candidate SM decomposed fields in the SO(10) multiplets in
TABLE I.
For reader's convenience, we also give the SU(5)
representation which includes the candidates in terms of  SM decomposed fields.
All four candidates are also included in 
${\bf 144} + \overline{\bf 144}$.
We comment that larger dimensional representations such as three- and
four-symmetric
tensors also include  candidate representations whose $N_A$ and $N_B$
are larger,
but there is no motivation to employ such fields.

\begin{table}
\caption{List of the fields whose $N_A$ and $N_B$ are both positive.
The definitions of $N_A$ and $N_B$ are given in the text.}
\begin{ruledtabular}
\begin{tabular}{ccccc}
& $N_A$ & $N_B$ & SO(10) & SU(5) \\ \hline
$({\bf 8},{\bf 2},1/2) + c.c.$ & $\frac{24}5$ & 24 &
${\bf 126} + \overline{\bf 126}$, $\bf 120$ & ${\bf 45},{\bf 50}$ \\
$({\bf 6},{\bf 1},1/3) + c.c.$ & $\frac{54}5$ & 6 &
${\bf 126} + \overline{\bf 126}$, $\bf 120$ & ${\bf 45}$ \\
$({\bf 6},{\bf 2},-1/6) + c.c.$ & $\frac{12}5$ & 36 & ${\bf 210}$ & ${\bf 40}$ \\
$({\bf 8},{\bf 1},0)$ & $6$ & 6 & ${\bf 210}$, $\bf 45$, $\bf 54$ & ${\bf 24},{\bf 75}$
\end{tabular}
\end{ruledtabular}
\end{table}

In the well motivated SO(10) representations,
the candidates for gauge unification at a higher scale are only four
decomposed fields as shown in the list.
Among them, the $SU(3)_c$ adjoint field $({\bf 8},{\bf 1},0)$
are already well known to increase $M_{H_C}$
in the context of SU(5) GUT models
\cite{Chkareuli:1998wi}.
However, the other three candidates have not been  discussed in the literature.
It is obvious that larger $N_B$ fields
can increase SO(10) breaking scale by a smaller hierarchy
between the breaking scale and the mass of the fields.
Therefore,
$({\bf 8},{\bf 2},1/2)$ and $({\bf 6},{\bf 2},-1/6)$
are more suitable candidates compared to others.
In fact,
if the particle contents are MSSM particles
plus $({\bf 8},{\bf 2},1/2)+ c.c$,
the 1-loop beta coefficients of $SU(3)_c$ and $SU(2)_L$
are same, $b_3 = b_2 = 9$, $b_1 = 57/5$.
If we adopt $({\bf 6},{\bf 2},-1/6)+c.c.$,
all three beta coefficients are same, $b_3 = b_2 = b_1 =7$ above the threshold.
Therefore,
it is possible
that the scale where the three gauge couplings meets (approximately)
is the mass scale of neither $({\bf 8},{\bf 2},1/2)$
nor $({\bf 6},{\bf 2},-1/6)$,
which are lighter
compared to the SO(10) breaking scale.
Actually, we have checked that there is a SO(10) breaking vacuum
where these candidate fields are light and no other decomposed field is light
by using a full expression of the Higgs spectrum \cite{Fukuyama:2004xs}.
We plot the gauge coupling running with $({\bf 8},{\bf 2},1/2)$ threshold
in FIG.1. We use 2-loop RGEs for our numerical calculation
and find the gauge coupling unification at the string scale.
Above the threshold of SO(10) breaking scale (possibly at the Planck scale)
the gauge couplings will blow up rapidly due to the large representations
such as ${\bf 126}+ \overline{\bf 126}$, however at that point the
theory is presumably described by string theory.

It is important
that the suppression of the dimension five operator
can be easily done
by such a simple assumption.
This mechanism holds for any
GUT model, independent of the detail of the model buildings.
For example, in the SU(5) model, the $\bf 45$ representation which
is used to fit fermion masses provides one of
the candidate fields.

\begin{figure}[t]
 \center
 \includegraphics[viewport = 30 24 240 230,width=5.8cm]{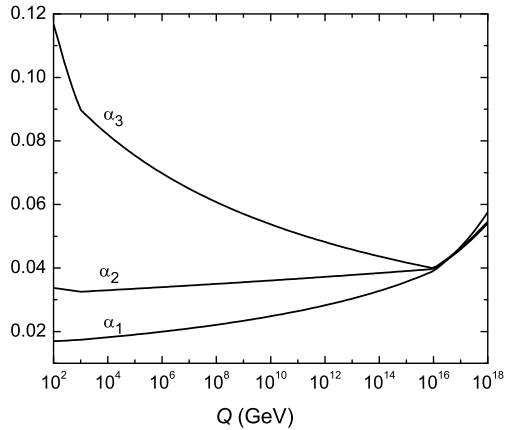}
 \caption{
Gauge coupling evolutions for MSSM with $({\bf 8},{\bf 2},1/2)$ threshold.
We choose $M_{({\bf 8},{\bf 2},1/2)}= 10^{16}$ GeV and $m_{\rm SUSY} = 1$ TeV.
Although the gauge symmetry does not recover at $10^{16}$ GeV,
the gauge couplings run unitedly above the threshold.
}
\vspace{-0.1mm}
\end{figure}

In models of the type discussed here,
if the mass scale of the candidate fields
are $10^{16}$ GeV, the colored Higgs fields
can be heavier than $10^{17}$ GeV
and the current nucleon decay bounds can be satisfied.
If $\tan\beta$ is large enough $\agt 20$,
the proton decay via dimension five operator
(such as $p\to K \bar\nu$)
can be observable in the megaton class detector.
However, the proton decay via the dimension six operator
(such as $p\to \pi e$) may not be observed since
it is suppressed by $M_X^4$,
while the dimension five nucleon decay is suppressed by $M_{H_C}^2$.

We comment that in the minimal SU(5) model, a heavy gluino
 lowers the $M_{H_C}$ and is therefore not admissible,
although heavy gluinos (heavier than other gauginos) are generic to
 minimal supergravity models.
However, in this SO(10) model, a heavy gluino is favored if we adopt
light $({\bf 8},{\bf 2},1/2)$.
Since $b_3 = b_2 < b_1$ (asymptotic non-freeness of U(1) is stronger
than the others)
is satisfied above the threshold, a heavier gluino threshold at low
energy makes $SU(3)_c$ and $SU(2)_L$
gauge couplings meet before $U(1)_Y$ coupling finally meets them at a
common unification.

It is important
that
if
$({\bf 8},{\bf 2},1/2)$
and/or $({\bf 6},{\bf 1},1/3)$
is much lighter than the SO(10) breaking scale,
a sizable flavor violation can be generated
since those fields originate from
$\overline{\bf 126}$ or $\bf 120$
which couple to fermions.
%
The couplings can be written as $q u^c \phi_{({\bf 8},{\bf 2},1/2)}
+q d^c \phi_{({\bf 8},{\bf 2},-1/2)}
+q q \phi_{(\bar{\bf 6},{\bf 1},-1/3)}
+u^c d^c \phi_{({\bf 6},{\bf 1},1/3)}.
$
Note that the $qq\phi$ coupling matrix is generation antisymmetric and
will give rise to specific flavor violation pattern. In general
 the flavor violating effects will depend on the origin
of the light fields.
For example, if $({\bf 8},{\bf 2},1/2)$ field
is light,
it can generate off-diagonal elements
for both left- and right-handed squark mass matrices.
On the other hand, if the light ${(\bar{\bf 6},{\bf 1},-1/3)}$
field comes form $\overline{\bf 126}$,
it can generate off-diagonal elements only for right-handed squarks.
In both cases we have found SO(10) breaking vacua where only each of
these above fields is light.
%
%

If both left- and right-handed squark mass matrices
have sizable off-diagonal elements,
the meson mixing via box diagram
is enhanced
and thus, it can have impact on the modification of the unitarity
triangle,
and $D$-$\bar D$ mixing \cite{Dutta:2006gq}.
If the flavor violation is generated from the symmetric couplings $f$,
$B_s$-$\bar B_s$ mixing phase can be enhanced
(when we  generate large atmospheric neutrino mixing)
rather than $K$-$\bar K$ and $B_d$-$\bar B_d$ mixings.
If they are generated form the antisymmetric couplings $h^\prime$,
modification of $B_d$-$\bar B_d$ phase and $D$-$\bar D$ mixing can be
sizable rather than the $B_s$-$\bar B_s$ phase.
This is because
the induced flavor non-universality is proportional to
$h^\prime h^{\prime\dagger}$
and
\begin{eqnarray}
h^\prime h^{\prime \dagger}
=
\left( \begin{array}{ccc}
          |b|^2+|c|^2 & -a^*b & -a^*c \\
         -ab^* & |c|^2+|a|^2 & -b^*c \\
         -ac^* & -bc^* & |a|^2+|b|^2
         \end{array}
  \right)
\end{eqnarray}
where $(h^\prime_{23},h^\prime_{13},h^\prime_{12}) = (a,b,c)$.
In a natural fit of the fermion mass, we have $a \gg b,c$.
As a result, we can distinguish the origins of $f$ and $h^\prime$
in the ongoing experiments.
In both cases, the 1-2 elements can be canceled in
the $Y_d$ diagonal basis
and then $K$-$\bar K$ mixing can be consistent with
the SM prediction.
However, we cannot cancel both $K$-$\bar K$ and $D$-$\bar D$
mixing amplitudes \cite{Dutta:2006gq}.
%
Therefore, this flavor violating threshold can affect the recently
measured $D$-$\bar D$ mixing.

We emphasize that none of the four candidates in the TABLE I
couples to leptons directly,
thus the enhancement of lepton flavor violations (LFV) such as $\tau \to
\mu\gamma$
is not favored
in the context of the proton decay suppression.
Actually,
if a decomposed field from the heavy Higgs fields which couples to
leptons is light,
it decreases either $M_{H_C}$ or $M_X$.
%
We comment that in the SO(10) models with simple fermion mass and mixing
fitting,
not much LFV 
is generated
from right-handed neutrino loops,
since the Dirac neutrino Yukawa coupling does not have large mixings.

It is also worthwhile to emphasize
that the LFV 
depends upon  which particles are light,
though it is not favored to increase $M_{H_C}$ 
For example, to realize type II seesaw,
$SU(2)_L$ triplet $({\bf 1},{\bf 3},1)$
needs to light
at the scale of intermediate scale $\alt 10^{14}$ GeV.
Then, the $f$ coupling can generate sizable
effects to the LFV. 
Also, if the $SU(2)_R$ breaking scale is smaller than the SO(10)
breaking scale, the
$SU(2)_R$ would-be-Goldstone Higgsino $({\bf 1},{\bf 1},-1)$
is light,
and then off-diagonal elements of the right-handed slepton
can be generated.
If the contribution arising from the $\bf 120$ coupling dominates,
the Br[$\tau \to e \gamma$] can be more enhanced compared to
Br[$\tau \to \mu\gamma$], which is an important prediction of this scenario.
Further analysis of quark and lepton flavor violations
depending on the SO(10) breaking vacua can be
found elsewhere.

Finally,
we comment on the $b$-$\tau$ unification
which is one of the important predictions of  GUT models.
In a natural fit, $\bar f_{33}$
is small and thus $b$-$\tau$ unification
is realized at the SO(10) breaking scale approximately
(up to 5\% $\sim m_\mu/m_\tau$).
{}From the numerical study of RGE evolution of Yukawa couplings, one finds that
the $b$-$\tau$ unification is satisfied
only when $\tan\beta \sim 2$ or $\sim 50$.
Now, $\bar f$ and $\bar h^\prime$ are small but
since these are multiplied by Higgs doublet mixings, the original
couplings $f_{33}$ and $h^\prime_{23}$
can be order 1 and, surely, the coupling to the $({\bf 8},{\bf 2},1/2)$
is not multiplied by the Higgs doublet mixings.
Then,
due to the $({\bf 8},{\bf 2},1/2)$ threshold, the
RGE evolution of $b$-$\tau$ Yukawa coupling can be modified,
and thus it is possible to realize the $b$-$\tau$ unification
even for a moderate $\tan\beta$.
In this situation, due to the large $f$ and $h$ couplings, the  induced
FCNCs can be also sizable.

In conclusion, we point out that
the lightest colored Higgsino mass
can be made heavy
when the fields listed in the TABLE I
are  light
compared to the SO(10) breaking scale and this pectrum is realizable in a
particular SO(10) breaking vacuum.
This will suppress the dimension five proton decay
which however still may be observable
in the next generation of 
detector.
We point out that the
constraints from nucleon decays
may imply sizable quark FCNCs
rather than leptonic FCNCs,
which can be tested in future. This can probe the detailed nature of GUT
theories.

The work of B. D. and Y. M. is supported in part by the DOE grant
DE-FG02-95ER40917. The work of R.N.M. is supported by the National
Science Foundation Grant No. PHY-0652363.

\end{document}